\documentclass[final,5p,twocolumn, review]{elsarticle}

\usepackage{amssymb}
\usepackage{amsmath}
\usepackage{graphicx}
\usepackage{booktabs}
\usepackage{multirow}
\usepackage{subfigure}
\usepackage{amssymb}
\usepackage{xcolor}
\usepackage[switch]{lineno}
\usepackage[utf8]{inputenc}
\usepackage[T1]{fontenc}
\usepackage[switch]{lineno}

\let\oldequation\equation
\let\oldendequation\endequation

\renewenvironment{equation}
  {\linenomathNonumbers\oldequation}
  {\oldendequation\endlinenomath}

 \journal{Energy Conversion and Management}

\begin{document}

\begin{frontmatter}



\title{A Graded Metamaterial for Broadband and High-capability Piezoelectric Energy Harvesting}


\author[inst1]{Bao Zhao\corref{mycorrespondingauthor}}
\cortext[mycorrespondingauthor]{Corresponding author}
\ead{bao.zhao@ibk.baug.ethz.ch}
\author[inst1]{Henrik R. Thomsen}
\author[inst2]{Jacopo M. De Ponti}
\author[inst3]{Emanuele Riva}
\author[inst4]{Bart Van Damme}
\author[inst4]{Andrea Bergamini}
\author[inst1]{Eleni Chatzi}
\author[inst1]{Andrea Colombi}

\affiliation[inst1]{organization={Department of Civil, Environmental, and Geomatic Engineering, ETH Zürich},
            city={Zürich},
            country={Switzerland}}

\affiliation[inst2]{organization={Department of Civil and Environmental Engineering, Politecnico di Milano},
            city={Milano},
            country={Italy}}

\affiliation[inst3]{organization={Department of Mechanical Engineering, Politecnico di Milano},
            city={Milano},
            country={Italy}}

\affiliation[inst4]{organization={Laboratory for Acoustics/Noise Control, Empa Materials Science and Technology},
            city={Dübendorf},
            country={Switzerland}}

\begin{abstract}
This work proposes a graded metamaterial-based energy harvester integrating the piezoelectric energy harvesting function targeting low-frequency ambient vibrations (<100 Hz). The harvester combines a graded metamaterial with beam-like resonators, piezoelectric patches, and a self-powered interface circuit for broadband and high-capability energy harvesting. Firstly, an integrated lumped parameter model is derived from both the mechanical and the electrical sides to determine the power performance of the proposed design. Secondly, thorough numerical simulations are carried out to optimise both the grading profile and wave field amplification, as well as to highlight the effects of spatial-frequency separation and the slow-wave phenomenon on energy harvesting performance and efficiency. Finally, experiments with realistic vibration sources validate the theoretical and numerical results from the mechanical and electrical sides. Particularly, the harvested power of the proposed design yields a five-fold increase with respect to conventional harvesting solutions based on single cantilever harvesters. Our results reveal that by bridging the advantages of graded metamaterials with the design targets of piezoelectric energy harvesting,  the proposed design shows significant potential for realizing self-powered Internet of Things devices.

\end{abstract}


\end{frontmatter}

\section{Introduction}
	\label{sec_intro}

\begin{figure}[!t]
		\centering
		\includegraphics[width=1\columnwidth,page=1]{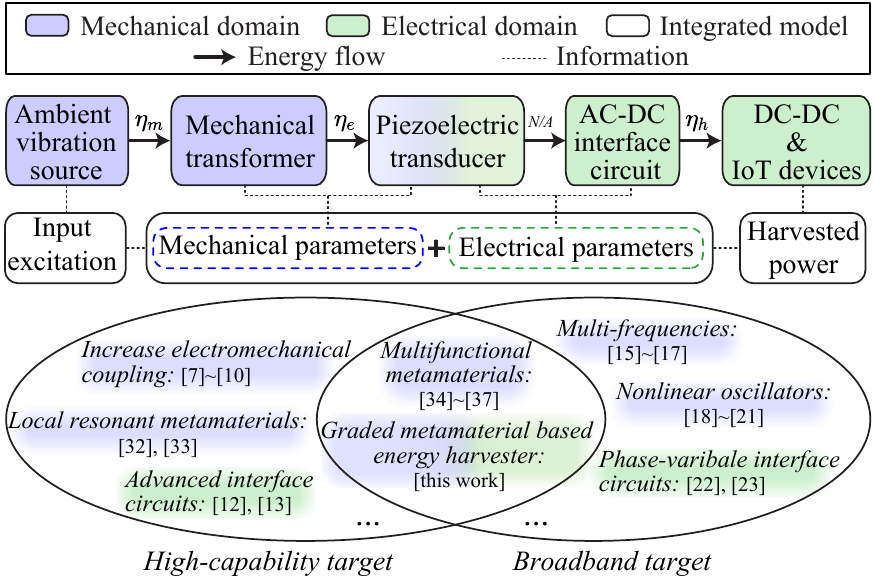}
		\caption{Block diagram of a piezoelectric energy harvesting (PEH) IoT system with its integrated model and design targets.}
		\label{fig_flow}
	\end{figure}

Energy harvesting has received considerable attention over the last two decades mainly in the context of transitioning toward the Internet of Things (IoT) architectures with milliwatt-level sensor nodes \cite{erturk2011piezoelectric}. The exploration of renewable energy sources not only relates to academic advances, but embraces significant social and economic values \cite{sharma2022review, huang2021effect, huang2022control}. Since vibration energy harvesting (VEH) leverages one of the most ubiquitous and accessible energy sources, this research field has tremendous potential to replace conventional, limited-life, chemical batteries for energy-efficient IoT devices.
Among different energy transduction methods \cite{elvin2013advances}, piezoelectric transduction is commonly used owing to its high power density and ease of integration in the design of compact energy harvesters. Through the so-called direct and reverse piezoelectric effect \cite{jaffe1958piezoelectric}, an external force applied on piezoelectric materials can be converted into an electrical voltage across the material's electrodes and vice versa. As the block diagram shown in Fig. \ref{fig_flow}, a piezoelectric energy harvesting (PEH) system can commonly be attained by attaching piezoelectric transducers on a mechanical transformer under the excitation of an ambient vibration source. The generated AC (alternating current) voltage can then be regulated with a power conditioning interface circuit into a DC (direct current) voltage \cite{erturk2011piezoelectric}. Irrespective of the specific mechanical design, transducer, and interface circuit chosen, the two primary design targets for PEH systems are:
	\begin{enumerate}
		\item \textit{The high-capability target}: to increase the harvested power at resonance;
		\item \textit{The broadband target}: to increase the off-resonance harvested power, i.e., to broaden the harvesting bandwidth.
	\end{enumerate}

Many studies have investigated these two targets for PEH systems from the mechanical and the electrical standpoints. To achieve the high-capability target (1), mechanical solutions include an increase in the number of active materials, or a decrease of the equivalent mechanical stiffness \cite{Ng2005,Cottone2012,wang2019integration,cha2019parameter}. All mechanical solutions yield an increased electromechanical coupling coefficient, which results in a stronger energy harvesting capability \cite{Shu2007}. Without altering the mechanical structure, advanced interface circuits \cite{zhao2020series,zhu2012theoretical} can also enhance the energy harvesting capability thanks to the increased impedance matching ability from the electrical side \cite{liao2018maximum}. For fulfilling the broadband design target, most research efforts stem from the mechanical engineering community with a variety of available options, namely combinations of multiple vibrators with different resonant frequencies \cite{xie2019energy,TadesseMultimodal,ErturkMultimodal} and introduction of nonlinear dynamics in the vibrator \cite{cai2022, wang2021nonlinear, qian2020bio, CottoneNonlinear2009}. Nevertheless, in strong coupling systems, the interface circuits can also contribute to the broadband target by additional electrically induced stiffness with phase-variable (PV) interface circuits \cite{lefeuvre2017analysis, zhao2020dual}. It is important to note that both mechanical and electrical designs play a synergistic role in the two design targets of PEH systems. The mechanical and electrical design integration is quickly becoming quintessential as piezoelectric harvesters are increasingly incorporated into metamaterial-based structures for enhanced energy harvesting \cite{hu2021acoustic, de2021graded, wang2021exploring}.

 Engineered materials can be mainly categorized into Bragg scattering phononic crystals \cite{sigalas1992elastic} and locally resonant metamaterials \cite{liu2000locally}. When aiming to harvest mechanical energy from low frequencies to fit the scenarios of IoT devices, the relatively long wavelength of low-frequency ambient vibrations renders the use of Bragg scattering structures inefficient. It is well known that locally resonant metamaterials  can surpass the size limitation of Bragg scattering systems, and generate sub-wavelength bandgaps leveraging local resonance mechanisms \cite{craster2012acoustic}. Sugino et al. \cite{sugino2016mechanism, sugino2018analysis} pointed out that the utility in using locally resonant metamaterials to enhance energy harvesting lies in their unique properties to slow down the propagation of elastic waves and focus the mechanical energy into the local resonators within or close to the frequency range of a bandgap. Therefore, the high-capability target of PEH can be naturally addressed using metamaterials. Li et al. \cite{li2017design} proposed a piezoelectric cantilever-based metamaterial for simultaneous vibration isolation and energy harvesting. Chen et al. \cite{chen2019metamaterial} also realized the same dual-function with a membrane-type metamaterial and further increased the output power with double-layer resonators.

 Despite the high-capability PEH by locally resonant metamaterials, the harvested power is noticeable only close to the bandgap frequency \cite{hu2021acoustic}, hence in a relatively narrow band. Therefore, research efforts have also investigated multifunctional metamaterials for tackling the broadband target \cite{hu2021acoustic}. Hwang and Arrieta \cite{hwang2018input} adopted nonlinear resonators and realized an input-independent metamaterial to broaden the energy harvesting bandwidth. Without resorting to nonlinear dynamics, De Ponti et al. \cite{de2020graded,de2020experimental,de2021enhanced} explored the rainbow trapping phenomena \cite{tsakmakidis2007trapped,jimenezacoustic, colombi2017enhanced} and realized graded metamaterials with simple and inherent broadband design. It can be seen that the research on enhanced energy harvesting with metamaterials mainly stems from the mechanical side. In contrast, the efforts from the electrical side for PEH with metamaterials are quite deficient. As shown in Fig. \ref{fig_flow}, considerations regarding AC-DC interface circuits and IoT devices are essential parts of holistically vibration-powered IoT devices. It still remains an open question how to transfer the advantages of metamaterials mentioned above into electromechanically integrated PEH systems and address the power consumption of IoT devices under low-frequency vibrations.

Building on the recent published graded metamaterials by De Ponti et al. \cite{de2020graded,de2020experimental,de2021enhanced}, we re-engineer and optimize a graded metamaterial-based energy harvester. For conciseness, this is referred to as "graded-harvester" hereafter. Different from previous results focusing on wave propagation under relatively high frequency ranges ($>$1 kHz) with absorbing boundary conditions, the proposed graded-harvester employs a clamped-free boundary condition to fit the practical PEH scenarios, which is capable to reach low frequencies and amplify the wave field, thus rendering operation under typical ambient vibration feasible ($<$100 Hz). Compared to the commonly used resistors as loads for PEH with metamaterials, the electrical part of the graded-harvester utilizes a self-powered interface circuit able to rectify the power produced by piezoelectric patches and provide a usable DC power supply for IoT devices. By combining the graded metamaterial and the AC-DC interface circuit through an integrated lumped model, the energy harvesting performance of the graded-harvester can be determined for the first time. The two design targets of PEH are thoroughly discussed with theoretical, numerical, and experimental analyses.

Section \ref{sec_intro} introduces the technological background and the proposed design. Section \ref{sec_theor} proposes an integrated model to calculate the harvested power and demonstrates the broadband energy harvesting ability of the design. Section \ref{sec_num} further discusses the wave propagation in the graded metamaterial for the high-capability energy harvesting target. Finally, Section \ref{sec_exp} shows experimental results for validating the design.

\begin{figure*}[!t]
		\centering
		\includegraphics[width=2\columnwidth,page=2]{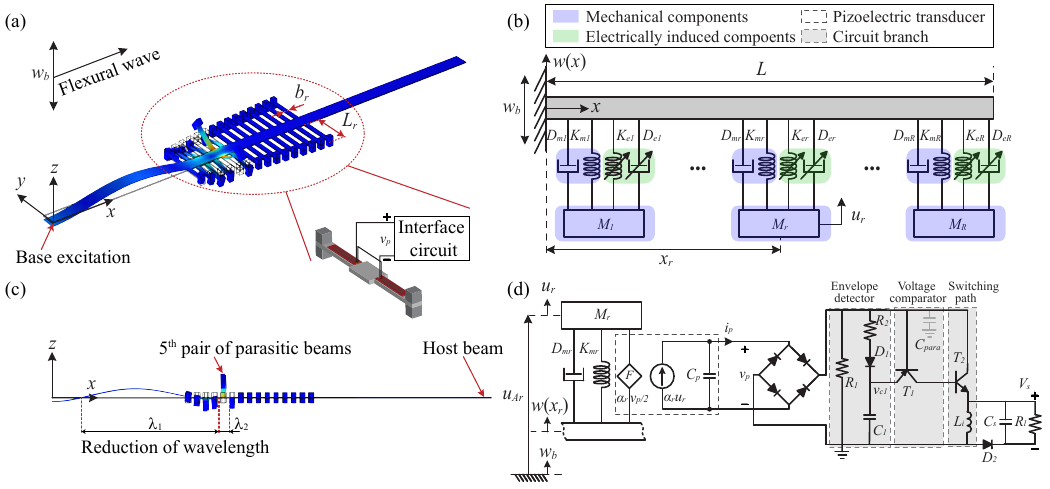}
		\caption{Illustrations of the graded metamaterial-based energy harvester and its theoretical model. (a) shows the flexural wave propagating in the graded metamaterial under base excitation in the $z$ direction. The enlarged view shows the configuration of piezoelectric parasitic beams.
		(b) shows the model schematic corresponding to (a). (c) illustrates the $xz$ cross-section of the structure depicted in (a). (d) displays the detailed equivalent parameter model of each piezoelectric parasitic beam.}
		\label{fig_model}
\end{figure*}

\section{Theoretical Analysis}
	\label{sec_theor}

In order to investigate the performance of energy harvesting systems, research efforts have been dedicated to theoretical \cite{Liang2012} and numerical methods \cite{de2020graded} to determine the harvested power and the bandwidth.  Unlike the wave propagation at high frequencies typical of acoustic metamaterials satisfying the assumptions of traveling waves in a long structure when compared to the wavelength \cite{craster2012acoustic}, a quasi-standing wave dominates the dynamic response of the finite structure under low-frequency vibrations \cite{sugino2016mechanism}. From the electrical side, without the assumption of pure resistive loads for energy harvesting \cite{de2020graded}, the equivalent impedance of the AC-DC interface circuit should be considered \cite{Liang2012}. In addition, the nonlinear coupling of interface circuits with mechanical structures \cite{wang2020new} often renders the numerical simulation difficult due to the intensive computation and memory requirements. For these reasons, a simplified integrated model combining mechanical and electrical lumped parameters is developed in this paper. The flow chart of the integrated model used in this section is shown in Fig. \ref{fig_flow}. The limited complexity of such a model, enables us to evaluate the energy harvesting performance of the graded-harvester with the chosen interface circuit.

\subsection{Graded Metamaterial and Circuit Topology}


We base our design on previous work by De Ponti et al. \cite{de2020experimental,de2020graded}, which exploited a graded metamaterial and slow waves to enhance the harvested power.
Figure \ref{fig_model} (a) shows the proposed design, which consists of a host cantilever beam and an array of parasitic beams with tip masses. The host beam of length $L$, width $b$, bending stiffness $EI$, and mass per length $m$ is excited by base excitation $w_b(t)$ in $z$ direction to generate a flexural wave traveling along the $x$ direction. $R$ pairs of parasitic beams of length $L_r$ and width $b_r$ are symmetrically distributed at positions $x_r$ along the host beam to minimize the torsional motion of the host beam. Identical tip masses and piezoelectric patches are attached to each pair of parasitic beams.

The graded design is achieved by gently increasing the length of each pair of parasitic beams. As shown in Fig. \ref{fig_model} (a) and (c), a wave generated by a sinusoidal base excitation undergoes a significant reduction of the wavelength along the host beam. At the location of the $5^{th}$ pair, the group velocity of the traveling wave generated by this example frequency vanishes. On the one hand, the low speed extends the interaction time between the parasitic beams with the traveling wave inside the graded array. On the other hand, it locally boosts the vibration amplitude to fulfill the high-capability energy harvesting target. At the same time, when compared against the non-graded metamaterial-based energy harvesters \cite{sugino2016mechanism}, the effective energy harvesting bandwidth can also be increased due to the spatial frequency separation, quintessential in rainbow devices \cite{tsakmakidis2007trapped,colombi2017enhanced}. By combining the contributions of different pairs of parasitic beams located at increasing distances, the broadband ability can also be achieved.

In order to transform the mechanical energy confined inside the metamaterial into electricity, piezoelectric patches are mounted on the parasitic beams. Conventionally, the bulk of research on metamaterial energy harvesting is restricted to the use of resistors as loads of the piezoelectric patches to measure an AC output power \cite{de2020experimental,alshaqaq2020graded}. However, this choice is not well suited to the DC power supply demand of IoT devices. Several studies have explored AC-DC interface circuits in PEH systems for this purpose. The Standard Energy Harvesting (SEH) circuit is the most commonly used, which only utilizes a bridge rectifier. In order to boost the energy harvesting capability, more advanced interface circuits \cite{lefeuvre2005piezoelectric,lefeuvre2006comparison,guyomar2005toward} have utilized $LC$ resonance for regulating the phase difference between piezoelectric voltage and current.In this paper, the Self-Powered Synchronized Electrical Charge Extraction (SP-SECE) \cite{zhu2012theoretical} is chosen as the interface circuit since it does not need any external driving signal for the switches and it is load-independent. As will be demonstrated later, we can partially decouple the nonlinear behavior of the electromechanical PEH system induced by the chosen interface circuit, which depends on different load and input amplitude conditions.

Fig. \ref{fig_model} (d) shows the circuit topology of SP-SECE. The electrical part of the piezoelectric transducer can be regarded as a clamped capacitor $C_p$ for each pair of piezoelectric patches in series and a parallel current source $i_{eq}$ whose value is proportional to the parasitic beam's tip velocity $\dot u_r$. By neglecting the leakage resistance, the voltage across the piezoelectric transducer can be represented by $v_p$. With the electromechanical force-voltage coupling coefficient $\alpha_r$, the reaction force $2F=\alpha_r v_p$ from the piezoelectric patch is evenly distributed at each parasitic resonator. Besides a full bridge rectifier, the SP-SECE circuit also features a buck-boost converter. Nevertheless, unlike the approach adopted for conventional buck-boost converters in power electronics \cite{sahu2004low}, which follow the Pulse Width Modulation (PWM) controlled method, the switching cycle of SP-SECE follows the mechanical vibration cycles at the current and velocity crossing zero point for self-powered synchronized switching actions. The switching actions are generated by three branches in Fig. \ref{fig_model} (d): the first is an envelope detector consisting of a discharge resistor $R_1$,  $D_1$, and $C_1$; the second is a voltage comparator by an off-the-shelf PNP transistor $T_1$, and the third is a switching path formed by a NPN transistor $T_2$ and an inductor $L_i$.

In this paper, the two piezoelectric patches of each pair of parasitic beams are connected in series with the interface circuit so it can easily overcome the voltage threshold induced by the bridge rectifier and the switching path  under low-amplitude vibrations. In order to evaluate the energy harvesting performance of the graded-harvester, an integrated model is built in the following section from the mechanical and electrical sides.

\subsection{Equivalent Lumped Parameters}

With reference to the flow chart of the integrated model in Fig. \ref{fig_flow}, a modal analysis method is implemented in order to derive the lumped parameters of the proposed design from the mechanical side, as shown in the schematic in Fig. \ref{fig_model} (b). The schematic consists of a host cantilever beam and periodically distributed parasitic resonators corresponding to the host beam and parasitic beams in Fig. \ref{fig_model} (a), respectively. Two identical parasitic beams of each pair are attached to the same position of the host beam, generating a joint resonance effect. In order to derive the lumped parameters of the resonators considering the additional stiffness introduced by piezoelectric patches, each parasitic beam can be represented as a partially covered piezoelectric cantilever beam with tip mass under a Single Degree Of Freedom (SDOF) system assumption \cite{hu2019modelling}, where the first flexural vibration mode is taken into account.

The SDOF model of a parasitic resonator for a piezoelectric parasitic beam in Fig. \ref{fig_model} (b) can be represented with the PEH model in Fig. \ref{fig_model} (d). Under given base excitation $w_b$ on the clamped side of the host beam, the relative transverse displacement at $x_r$ of the host beam can be regarded as $w(x_r)$. The absolute displacement of the SDOF parasitic resonator at $x_r$ is $u_{Ar}(t)=w_b(t)+w(x_r,t)+u_r(t)$, as shown in Fig. \ref{fig_model} (d). Therefore, a parasitic beam with piezoelectric patches can be represented with a SDOF energy harvester under base excitation with acceleration  $\ddot{w}_{abs}=\ddot{w}_{b}+\ddot{w}$. The equivalent mass $M_r$, stiffness $K_{mr}$, damping coefficient $D_{mr}$, and the equivalent force-voltage coupling coefficient $\alpha_r$ can be formulated as:
\begin{equation}
\begin{aligned}
&M_r=\frac{1}{\psi_{r}^{2}\left(L_{r}\right)} ; \quad K_{mr}=\frac{\omega_{r}^{2}}{\psi_{r}^{2}\left(L_{r}\right)} ; \\
&D_{mr}=\frac{2 \zeta_{r} \omega_{r}}{\psi_{r}^{2}\left(L_{r}\right)} ;\quad \alpha_r=\frac{\beta_{r}}{\psi_{r}\left(L_{r}\right)}
\end{aligned}
\label{eq_1}
\end{equation}
where $\psi_r$ is the first order mode shape of the beam section without piezoelectric layer \cite{hu2019modelling} in the $r^{th}$ pair of parasitic beams. $\zeta_r$, $\beta_r$ and  $\omega_r$ describe the damping ratio, the coupling related factor and the first order resonance frequency of the $r^{th}$ beams, respectively.

	\begin{figure}[!t]
		\centering
		\includegraphics[width=0.95\columnwidth,page=3]{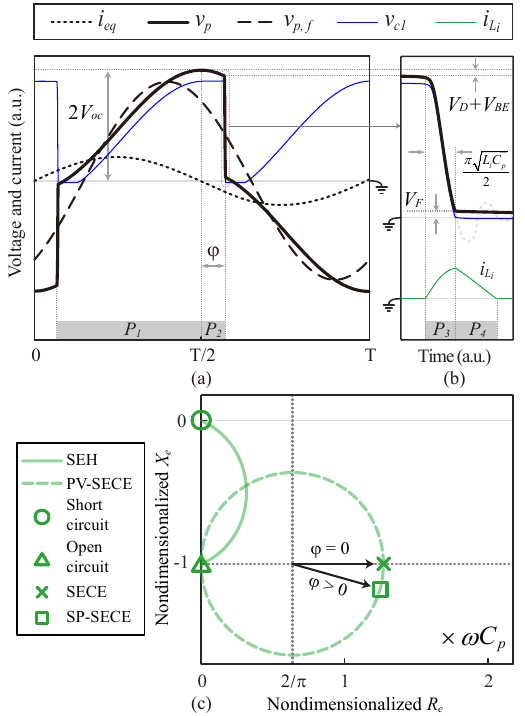}
		\caption{The waveform and equivalent impedance of SP-SECE interface circuit.
		(a) Operation waveform of SE-SECE. (b) $L_i$-$C_p$ resonance switching phase and the freewheeling phase. (c)  Nondimensionalized equivalent impedance of SEH, PV-SECE, SECE and SECE.}
		\label{fig_sece}
	\end{figure}

The first-order resonance frequency $\omega_r$ can be determined by solving the characteristic equation of the $r^{th}$ pair of parasitic beams. Along with the boundary conditions, the mode shape $\psi_r$ and the coupling related factor $\beta_r$ are computed. It should be noted that the derived lumped parameters are effective only at the beam's tip. The reaction force at the root of the beam derived from the lumped parameters is slightly different from that derived from the analytical model \cite{hu2019modelling}. However, by properly choosing a relatively large tip mass, these parameters are still valid for calculating the reaction force at the root of the beam \cite{hu2021comb}.


Additionally, we include the electrical side considering the lumped parameters of the interface circuit as shown in Fig. \ref{fig_flow}. We employ a general impedance analysis method \cite{Liang2012} to determine the lumped parameters, which relies on the computation of the first order electrically induced damping and stiffness coefficients: $D_e$ and $K_e$.
Figure \ref{fig_sece} (a) shows the waveform of the SP-SECE interface circuit, including its piezoelectric voltage $v_p$, current waveform $i_{eq}$, and the voltage $v_{cl}$ of $C_1$ (also the emitter voltage of $T_1$). In Fig. \ref{fig_sece}, $\varphi \in [0,\pi /2]$ describes the positive switching phase lag and $V_{oc}$ is the nominal open-circuit voltage depending on the relative displacement amplitude of a parasitic resonator.
The operation of SP-SECE consists of four phases from $P_1$ to $P_4$ as shown in Fig. \ref{fig_sece} (a) and (b):
	\begin{enumerate}
		\item $P_1$: an open circuit phase between the end of the last switching phase $P_3$ and the current crossing zero point;
		\item $P_2$: a switching delay phase, denoted by phase lag $\varphi$, due to the conduction of the switching path;
		\item $P_3$: a switching phase to transfer the charge from $C_p$ to $L_i$;
		\item $P_4$: a freewheeling phase to transfer the energy in $L_i$ into the storage capacitor $C_s$ through the freewheeling diode $D_2$.
	\end{enumerate}

Based on these four steps, the equivalent impedance $Z_e$ of SP-SECE interface circuit can be derived with impedance analysis (detailed in Appendix A). The electrically induced damping $D_{er}$ and stiffness $K_{er}$ of each parasitic resonator in the $r^{th}$ pair can be derived based on the electromechanical analogy in this coupling system, namely:
    \begin{equation}
    D_{er}=\alpha_r^{2} R_{e}/2, \quad K_{er}=-\alpha_r^{2} \omega X_{e}/2.
    \end{equation}
When the system operates at frequency $\omega$, the real part $R_e$ and the imaginary part $X_e$ of $Z_e$ form functions of $\varphi$, whose relationship is illustrated in the two-dimensional impedance plane depicted in Fig. \ref{fig_sece} (b). The figure shows the equivalent impedance of SECE (special case of PV-SECE \cite{lefeuvre2017analysis} when $\varphi =0 $) and SP-SECE (special case of PV-SECE when $\varphi >0 $). Compared with SEH, the real part $R_e$ of SECE or SP-SECE is larger, which means it has a stronger impedance matching ability to achieve a higher power output. By assuming $V_{oc}>10(V_D+V_{BE})$, the difference of $R_e$ between SP-SECE and SECE is less than 5\%. Therefore, we can approximate the equivalent impedance of SP-SECE by that of SECE and remove the dependency on input amplitude under the relatively large $V_{oc}$ condition. Consequently, different from the SEH case, the $D_{er}$ and $K_{er}$ of SP-SECE are independent of the load and input amplitude conditions, which decouple the system with invariant electrically induced components for the proposed metamaterial for energy harvesting.

	\subsection{Integrated Model}

As shown in Fig. \ref{fig_flow}, by combining the lumped parameters determined mechanically and electrically, each piezoelectric parasitic beam can be represented by a SDOF parasitic resonator with the SP-SECE interface circuit. Therefore, the governing equation for each parasitic resonator can be formulated as:
    \begin{equation}
    \small
        {M_r}{{\ddot u}_{r}}(t) + {D_r}{{\dot u}_r}(t) + {K_r}{u_r}(t) =  - {M_r}\left( {\frac{{{\partial ^2}w({x_r},t)}}{{\partial {t^2}}} + {{\ddot w}_b}(t)} \right),
        \label{eq:parasiticbeam}
    \end{equation}
where $K_r=K_{mr}+K_{er}$ and $D_r=D_{mr}+D_{er}$ contain the combined effect of the mechanical and electrical induced stiffness and damping, respectively. As stated previously, the electrically induced components are determined by the chosen interface circuit. For SP-SECE, $D_{er}$ is positive, which means it absorbs mechanical energy and converts it into electricity. In order to evaluate the energy harvesting performance of the proposed design, the displacement amplitude $U_{r}$ of the $r^{th}$ pair parasitic resonators shall be determined.

By adding the reaction forces of each pair of parasitic resonators onto the host beam, the governing equation of the host beam can be expressed as:
    \begin{equation}
    \begin{aligned}
    &EI \frac{\partial^{4} w}{\partial x^{4}}+m \frac{\partial^{2} w}{\partial t^{2}}=-m \ddot{w}_{b}(t)+ \\
    &2\sum_{r=1}^{R} K_{r} u_{r}(t) \delta\left(x-x_{r}\right)+2\sum_{r=1}^{R} D_{r} \dot{u}_{r}(t) \delta\left(x-x_{r}\right),
    \end{aligned}
    \label{eq:hostbeam}
    \end{equation}
where $\delta$ represents the Dirac function. The integrated model of the graded-harvester is then represented by Eq. \ref{eq:parasiticbeam} and  Eq. \ref{eq:hostbeam}.

    $U_{r}$ can be solved by modal analysis method with the electrically induced components from the SP-SECE interface circuit (detailed in Appendix B). Furthermore, the magnitude of the piezoelectric current of the $r^{th}$ pair of parasitic resonators in series can be expressed as \cite{Liang2012}:
    \begin{equation}
    I_{eqr} = \alpha _r \omega  U_{r}.
    \end{equation}
    Taking into account the rectifier loss, switching loss, and freewheeling loss \cite{chen2019revisit} of the SP-SECE interface circuit under the relatively large $V_{oc}$ assumption, the harvested power for each pair of parasitic resonators can be formulated as:
    \begin{equation}
    {P_{r}} = {\frac{{I_{eqr}^2{E_h}}}{{2\Delta E}}} {R_e}=\eta _h {\frac{I_{eqr}^2 }{2}} {R_e} ,
    \label{eq:ph}
    \end{equation}
    where $\Delta E$ and $E_h$ represent the total extracted energy and the harvested energy in one cycle. The ratio between $\Delta E$ and $E_h$ is denoted as the harvesting efficiency $\eta_h$:
    \begin{equation}
    \eta _h=\frac{E_h}{\varDelta E}=\frac{\left| \gamma \right|\tilde{V}_s}{\tilde{V}_s+\tilde{V}_D}\left( 1-\tilde{V}_F \right),
    \label{eq:eff_h}
    \end{equation}
    where $\gamma$, $\tilde{V}_s$, $\tilde{V}_D$, and $\tilde{V}_F$ describe the flipping factor of the SP-SECE interface circuit, the $V_{oc}$ normalized $V_s$, $V_D$ and $V_F$, respectively. Consequently, the harvested power is a percentage of the extracted power. All the parameters in Eq. \ref{eq:ph} are constant under the specific base excitation condition, except the harvesting efficiency $\eta_h$. Strictly speaking, the harvested power also changes with the load condition, which influences $\tilde{V}_s$. However, this dependency can be significantly weakened after the conduction of the freewheeling diode $D_2$ \cite{chen2019revisit}.

    As a case study, the theoretical harvested power, computed using the integrated model described above, for different pairs of parasitic beams is plotted as a gray line in Fig. \ref{fig_amp} with the parameters from Table \ref{para}. For the harvested power in single resonant energy harvesters \cite{Liang2012}, the bandwidth usually refers to the $-3$ dB bandwidth. For the graded-harvester, there is no standard criterion to define the harvested power bandwidth. However, it can be seen that inside the graded area, there are multiple harvested power peaks corresponding to the resonance frequencies of different pairs of parasitic beams in the range from 60 Hz to 160 Hz. Therefore, the broadband energy harvesting ability of the graded-harvester is guaranteed.

     \begin{figure*}[!t]
		\centering
		\includegraphics[width=2\columnwidth,page=4]{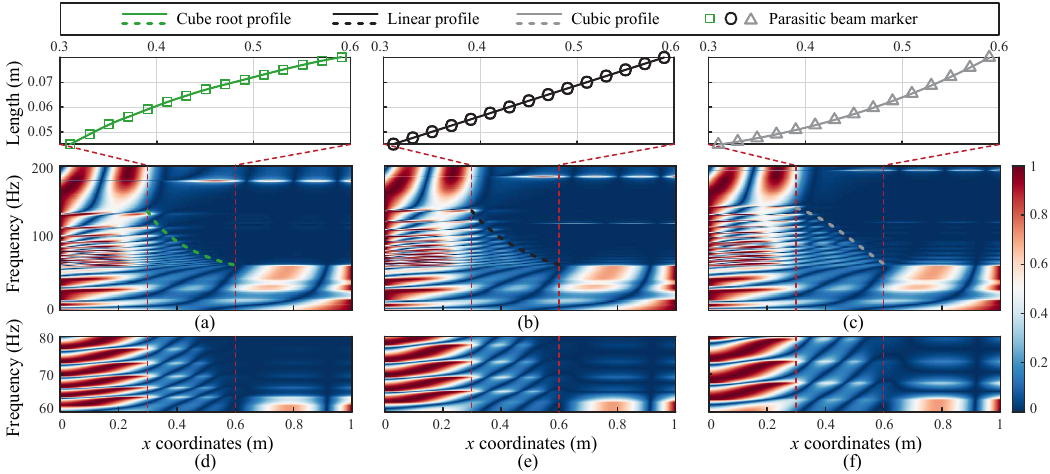}
		\caption{The different grading profiles and their numerical spatial-frequency analysis normalized at each frequency. (a) Cubic root profile. (b) Linear profile. (c) Cubic profile. (d), (e), and (f) show the 60-80 Hz enlarged view of (a), (b), and (c), respectively.}
		\label{fig_profile}
	\end{figure*}

	\section{Numerical Analysis}
	\label{sec_num}


    The grading design of the metamaterial not only broadens the energy harvesting bandwidth. It also affects the dispersion relationship and the wave field propagation, which further influences the energy harvesting ability of the designed graded-harvester. This section investigates wave propagation within the graded metamaterial using numerical simulations (FEM) and demonstrates the broadband and high-capability energy harvesting characteristic of the graded-harvester.

	\subsection{Grading Profile}
    In order to evaluate the harvesting ability in the bandgap range, it is necessary to investigate the role of different grading profiles. As shown in Fig. \ref{fig_profile}, the spatial-frequency analyses of three grading profiles for the parasitic beams are simulated in the frequency domain with COMSOL Multiphysics. The lengths of different pairs of parasitic beams distributed at the same positions on the host beam follow the cube-root profile, linear profile, and cubic profile, respectively. The three different grading profiles can be expressed as a function of the positions of different parasitic beams:
    \begin{equation}
    L_r=ax_{r}^{p}+b,
    \end{equation}
    where $p=1/3, 1, 3$ represents the cube-root, linear, and cubic profiles, respectively. By fixing the length of parasitic beam pair $L_1$ and $L_{15}$, parameter $a$ and $b$ can be determined. As shown in Fig. \ref{fig_profile}, the cubic-root, linear, and cubic profiles exhibit the same bandgap ranges but different spatial-frequency separation curves\footnote{It should be noted that the grading discussed here concerns the lengths rather than the resonance frequencies of parasitic beams determined by Eq. \ref{eq_1}. Therefore the final spatial-frequency separation curves indicated with dashed lines in Fig. \ref{fig_profile} (a), (b), and (c) do not have the same shapes corresponding to their grading profile functions.} where the wave propagation vanishes, as indicated with the green dashed line, black dashed line, and gray dashed line, respectively.  Compared with the linear or cubic profiles as shown in Fig. \ref{fig_profile} (e) and (f), the cubic-root one shown in Fig. \ref{fig_profile} (d) enables more localized modes at the beginning of the bandgap range due to the increased number of longer parasitic beams \cite{alshaqaq2020graded}. This creates a stronger bandgap and suggests a higher energy density, ideal for boosting energy harvesting, and it is thus chosen for the proposed design.

    \begin{figure}[!t]
		\centering
		\includegraphics[width=1\columnwidth,page=5]{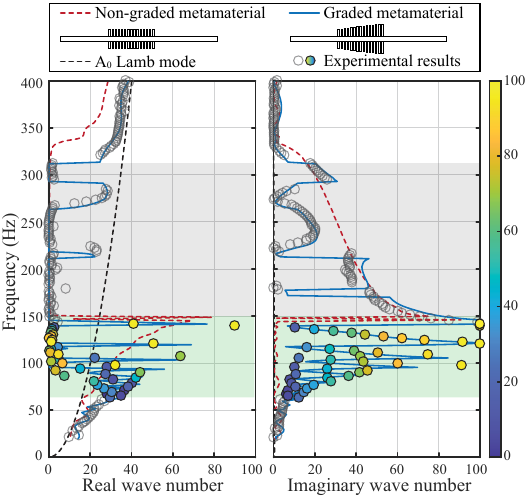}
		\caption{The dispersion relationship of the graded-harvester. The curves represent the numerical dispersion relationships of the $A_0$ Lamb mode (black dashed lines) of the host beam, the non-graded (red dashed lines) metamaterial, and the graded (blue lines) metamaterial, respectively. For experimental wave numbers of the graded metamaterial, the wave that propagates outside the grading frequency range is indicated by gray circles; the wave that propagates within the grading frequency range is indicated by color circles. In the grading frequency range, the real and imaginary parts of each complex wave number are labeled with the same color scale with respect to the amplitude of the imaginary wave number.}
		\label{fig_dispersion}
	\end{figure}

	\begin{figure*}[!t]
		\centering
		\includegraphics[width=2\columnwidth,page=6]{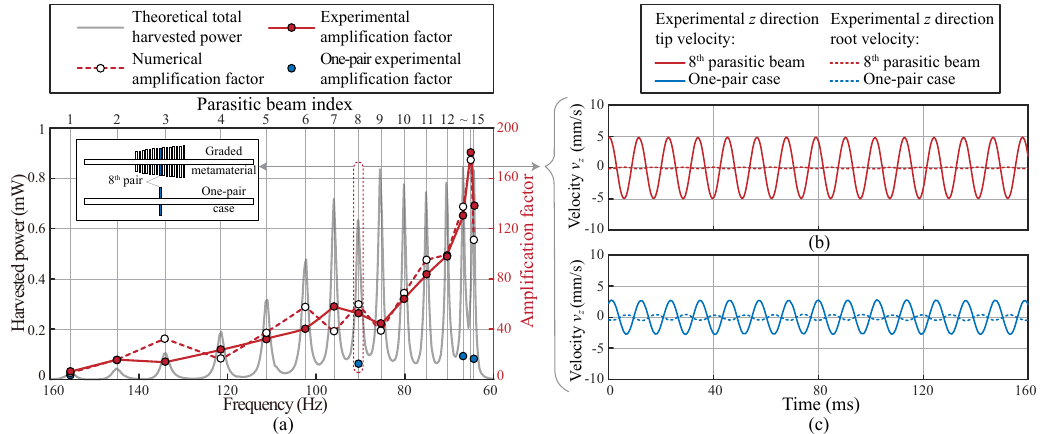}
		\caption{The harvested power and amplification factor of the graded-harvester. (a) shows the harvested power and amplification factor of the graded-harvester, the frequency axis is plotted in a descending order, according to the grading frequency distribution.  (b) shows the experimental velocity of the $8^{th}$ parasitic beam. (c) shows the experimental velocity of the one-pair case corresponding to the $8^{th}$ parasitic beam.}
		\label{fig_amp}
	\end{figure*}

	 \begin{figure}[!t]
		\centering
		\includegraphics[width=1\columnwidth,page=7]{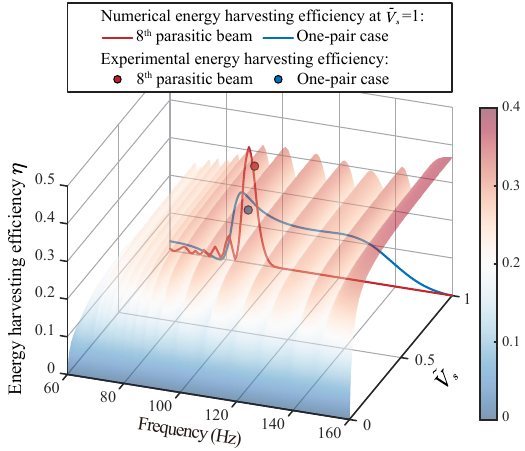}
		\caption{The energy harvesting efficiency $\eta$ of the graded-harvester. The semitransparent surface represents the envelope of energy harvesting efficiencies from different parasitic beams of the graded-harvester. }
		\label{fig_efficiency}
	\end{figure}

	\subsection{Numerical Results}
    The dispersion relationship provides a general and fundamental description of the wave propagation characteristics of linear metamaterials. It leads to a better understanding of how the propagating wave could enable higher energy harvesting ability spanning the grading frequency range \cite{colombi2017enhanced}. Based, therefore, on the cubic-root grading profile, we discuss wave propagation in the graded metamaterial by dispersion analysis. Due to the out-of-plane base excitation, traveling flexural waves ($A_0$) in the host beam are dominant versus longitudinal and torsional waves, or higher-order Lamb waves. Limited by the length of the host beam, the commonly used two-dimensional Fourier transform (2D-FFT) \cite{van2011time} can not give a high-resolution dispersion relationship at low-frequency vibrations. As an alternative, we further herein adopt the Inhomogeneous Wave Correlation (IWC) method \cite{van2018measuring}, which allows for accurate characterization of the dispersion relationship over short measurement distances. The wave number is calculated by the maximized correlation between a theoretical inhomogeneous running wave and the spatial response from measurements or simulations.


Figure \ref{fig_dispersion} shows the numerical dispersion curves of the $A_0$ Lamb mode, a non-graded metamaterial, and the graded metamaterial from frequency domain analysis with COMSOL Multiphysics. Compared to the graded metamaterial, the non-graded metamaterial simulated here represents the commonly used periodic design  \cite{sugino2016mechanism} with the same parasitic beam length equal to $L_1$. The different structures are excited with a constant acceleration field. The numerical dispersion curves are computed via the IWC method, using the simulated frequency-domain velocity responses spaced at 1mm intervals across the host beam. For the dispersion curves of the graded metamaterial, shown in blue lines, the wave propagates as a $A_0$ Lamb mode with no dissipation outside the bandgap range. Inside the bandgap range indicated by the gray shadow (starts from the resonant frequency of the $1^{st}$ pair of parasitic beams at 154 Hz to 317 Hz), most of the real wave numbers are zero, which means there is no wave propagation in this frequency range. Furthermore, the positive imaginary wave numbers indicate the exponential decay of traveling waves due to the local resonant effect \cite{fano}.

Besides the bandgap creation described above, the graded design introduces an additional gap, which is indicated via a green shadow (the grading frequency range from the resonant frequency of the $15^{th}$ pair of parasitic beams at 66 Hz to 154 Hz). For the real wave numbers, the non-vanishing values become larger as the frequency increases. Rather than propagating through the metamaterial as in the non-graded case, the waves slow down at the positions of local resonators with group velocity and wavelength reduction. This slow-wave phenomenon can also be seen from the imaginary wave numbers. The imaginary wave numbers are more prominent in the graded range than in the non-graded case, which also explains the exponential decay of wave propagation with the graded design. The slow waves enable longer interaction with the local resonators, which leads to the amplification of the wave fields.

We introduce the amplification factor of the wave field in local resonators as a metric of the energy transfer capability from the metamaterials to local resonators. We firstly define the relative velocity of a local resonator as the velocity difference between its tip and root positions of a parasitic beam: $\dot u_r^{\prime}=\dot{u}_r-\dot{w}\left( x_r \right)$. Then the amplification factor can be expressed  as the ratio of the relative velocity amplitude of a local resonator to the velocity amplitude at its root position on the host beam:
$\dot U_r^{\prime}/\dot{W}\left( x_r \right)$. Figure \ref{fig_amp} (a) shows the simulated amplification factors of each pair of parasitic beams at their resonance frequencies using COMSOL time domain simulations. It is observed that the amplification factor of the graded metamaterial increases with the length of the parasitic beams, except for a drop at the $15^{th}$ parasitic beam due to the wave leakage at the beginning of the bandgap and the lack of additional resonators. The trend of the amplification curve follows that of the harvested power curve in the graded frequency range. This amplification of the wave field shows the potential to harvest mechanical energy with the graded metamaterial efficiently.

With the wave field amplification by the graded metamaterial, we further calculate the energy harvesting efficiency of the graded-harvester. From the perspective of energy flow  \cite{zhao2020series,yang2017efficiency} shown in Fig. \ref{fig_flow}, the energy transformation process of the graded-harvester contains three steps:
	\begin{enumerate}
		\item The mechanical energy inside the graded metamaterial is efficiently transferred to local resonators as mechanical transformers. The efficiency of this step is defined as $\eta_{m}$;
		\item The mechanical energy in local resonators is converted into the electrical extracted energy with piezoelectric transducers. The efficiency of this step is defined as $\eta_{e}$;
		\item  By considering the piezoelectric transducer and the interface circuit as a whole, the electrical extracted energy is further converted into the net energy harvesting energy in the storage considering different losses. The efficiency of this step is $\eta_{h}$ in Eq. \ref{eq:eff_h}.
	\end{enumerate}
In order to determine the total energy harvesting efficiency $\eta=\eta_{m} \eta_e \eta_h$, we firstly simulate $\eta_{m}$ of each pair of parasitic beams. From energy conservation, the work done by the external load is finally consumed by the damping effect of the graded-harvester. We use an isotropic mechanical damping coefficient 0.001 for the graded metamaterial and the electrically induced damping $D_{er}$ for each pair of parasitic beams. Therefore, $\eta_{m}$ can be simulated by the ratio between the dissipation energy of each pair of parasitic beams and the total dissipation energy of the graded-harvester. Then, the transferred mechanical energy is converted into electrical energy with the efficiency $\eta_{e}=D_{er}/ D_r$. By considering the electrical losses in the interface circuit, the total numerical energy harvesting efficiency $\eta$ is shown in Fig. \ref{fig_efficiency}. The efficiency of the $8^{th}$ parasitic beam at $\tilde{V}_s=1$ reaches the peak value at its resonance frequency, after which the efficiency drops to zero due to the spatial frequency separation mentioned above. Compared to the one-pair case indicated with the blue line, the $\eta$ of the $8^{th}$ parasitic beam increased 160\% due to the wave field amplification by the graded metamaterial. This kind of relay of different parasitic beams forms the envelope surface of the total efficiency $\eta$, which maintains a high level in the grading frequency range and serves the high-capability energy harvesting target.

\section{Experiment}
\label{sec_exp}

Finally, both the broadband and high-capability energy harvesting of the designed graded-harvester are validated experimentally. From the mechanical side, the broadband energy harvesting induced by the graded metamaterial design is investigated via transmissibility and spatial-frequency analysis. Additionally, the high-capability energy harvesting is demonstrated by means of the experimental dispersion relationship and the amplification factor. From the electrical side, the practical energy harvesting performance of the graded-harvester is evaluated by comparison to a beam with only one pair of parasitic beams fixed to the host beam.

	\begin{figure}[!t]
		\centering
		\includegraphics[width=1\columnwidth,page=8]{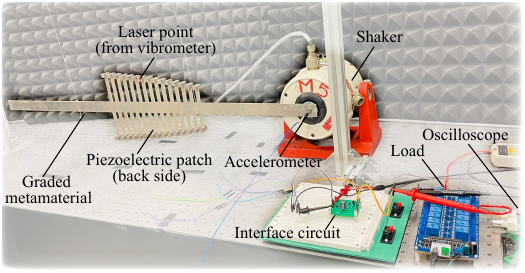}
		\caption{Experimental setup.}
		\label{fig_setup}
	\end{figure}

    \subsection{Setup}

    \begin{table}[t!]
     \footnotesize
    \caption {Parameters in Experiment} \label{para}
    \begin{center}
    \setlength{\tabcolsep}{2mm}{
    \begin{tabular}{llll}
    \toprule
    \multicolumn{4}{c}{\textbf{Graded Metamaterial based Energy Harvester}}     \\
    \midrule
    \multicolumn{4}{l}{\textbf{Host Beam}}                                \\
    Size   & 1000$\times$30$\times$2 (mm$^3$) & Material  & Aluminum
    \vspace{0.05in} \\
    \multicolumn{4}{l}{\textbf{Parasitic Beam}}                          \\
    Size   & $L_r \times$10$\times$2 (mm$^3$)  & Material  & Aluminum   \\
    \multicolumn{4}{l}{ $L_r=21.6\sqrt[3]x_r-101.3$ (mm), \quad $x_r=310:20:590$  (mm)}
    \vspace{0.05in} \\
    \multicolumn{4}{l}{\textbf{Tip Mass}}                                 \\
    Size   & 10$\times$10$\times$10 (mm$^3$)  & Material & Steel
    \vspace{0.05in} \\
    \multicolumn{4}{l}{\textbf{Piezoelectric Patch}}                      \\
    Size   & 40$\times$7$\times$0.8 (mm$^3$)    & Material & PZT            \\
    $d_{31}$  & -60 pC/N                        & $C_p$ & 19.6 nF
    \vspace{0.05in} \\
    \multicolumn{4}{l}{\textbf{Interface Circuit}}               \\
    Diodes  & SS14 ($V_D=0.5$ V)     & $L_i$    & 10 mH         \\
    MOSFETs & XN4601  ($V_{BE}=0.5$ V) & $C_s$    & 4.7 $\mu$F    \\
    Rectifier & MB6S ($V_{F}=1$ V)   &  $\gamma$ &   -0.63\\
    \bottomrule
    \end{tabular}}
    \end{center}
    \end{table}

    Figure \ref{fig_setup} shows the experimental setup. The graded metamaterial was fabricated out of one aluminum plate by water-jet cutting. Two identical tip masses and one piezoelectric patch are attached to each parasitic beam.
    The graded metamaterial is then hung vertically by two ropes to maintain stability.
    One end of the graded metamaterial with shorter parasitic beams is clamped to a shaker (LDS V406) for base excitation resulting in a clamped-free boundary condition. The two piezoelectric patches of one pair of parasitic beams are connected in series. Because the base excitation is symmetric, the antisymmetric modes of the pair of parasitic beams are not excited \cite{de2021enhanced}. This means each pair is free from out-of-phase vibration and charge cancellation. Therefore, it is feasible to configure two piezoelectric patches of one pair of parasitic beams in series. The piezoelectric voltage is connected to the SP-SECE interface circuit for energy harvesting. The parameters of the graded-harvester are shown in Tab. \ref{para}. These mechanical and electrical parameters were chosen to realize the energy harvesting function of the graded-harvester, considering mainly off-the-shelf, easily customizable (printed circuit board), and inexpensive components. The shaker is triggered simultaneously with a Polytec 3D scanning laser Doppler vibrometer (SLDV), which records the 3D velocity field at the center-axial points on the host beam with $5$ mm spacing and the tip points of parasitic beams in the time domain through repeated acquisitions.

    \subsection{Experimental Results}

    \begin{figure}[!t]
		\centering
		\includegraphics[width=1\columnwidth,page=9]{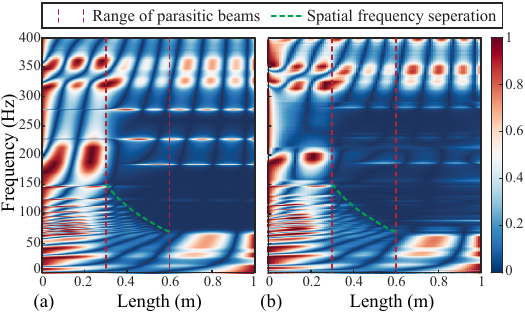}
		\caption{Comparison of spatial-frequency analyses.  (a) and (b) show the numerical and experimental spatial-frequency analyses of the graded metamaterial. The spatial-frequency separation is shown with the green dashed line. The graded area is indicated with two vertical red dashed lines. }
		\label{fig_spatialfre}
	\end{figure}

	 \begin{figure}[!t]
		\centering
		\includegraphics[width=1\columnwidth,page=10]{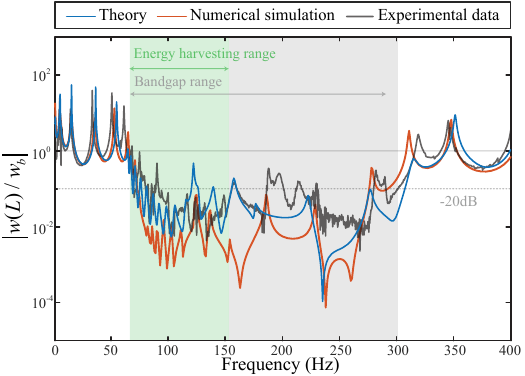}
		\caption{Comparison of transmissibilities.  The three curves show the theoretical, numerical, and experimental tip transmissibility, respectively. }
		\label{fig_bandgap}
	\end{figure}

		\begin{figure*}[!t]
		\centering
		\includegraphics[width=2\columnwidth,page=11]{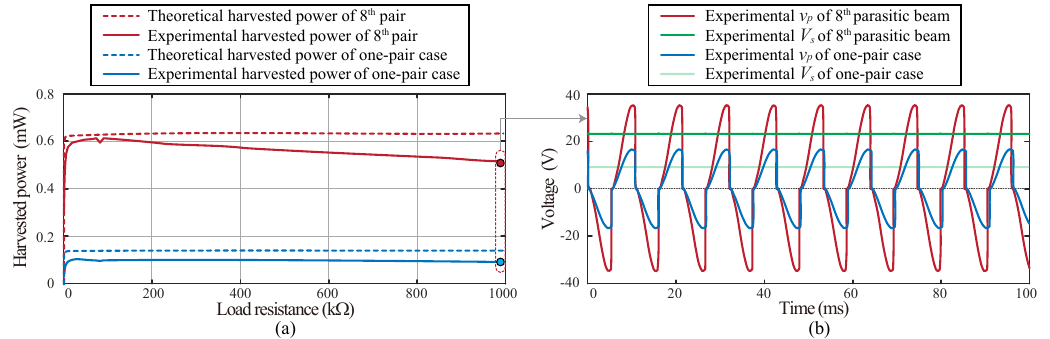}
		\caption{The harvested power of the $8^{th}$ pair parasitic beams and its corresponding one-pair case  under $0.1$N harmonic based excitation at their own resonance frequencies. (a) shows the theoretical and experimental harvested power of $8^{th}$ pair parasitic beams and its one-pair case. (b) shows the piezoelectric voltage $v_p$ and the storage voltage $V_s$ under 1M$\Omega$ load condition. }
		\label{fig_power}
	\end{figure*}

    Figure \ref{fig_spatialfre} (a) and (b) show the spatial-frequency separation of the graded metamaterial from numerical simulation and experimental data. The numerical simulation setup is identical to the procedure discussed in Sec. \ref{sec_num}. In the experiment, the beam was excited with a broadband white noise signal. The experimental spatial-frequency was obtained from the Fourier transformed time-domain velocity responses of the center-axial points on the host beam. The bandgap range of the experimental data shows good agreement with the numerical simulation. Inside the graded bandgap area, the velocity amplitude and the wavelength decrease in agreement with Fig. \ref{fig_spatialfre} (a), until it vanishes as shown with the green dashed lines in correspondence with the positions of the local parasitic beams. As stated before, this spatial-frequency separation  enables the broadband energy harvesting target. From the bandgap perspective, Fig. \ref{fig_bandgap} shows the tip transmissibility under open circuit condition of the graded metamaterial. The results displayed are from the theoretical model derived from Sec. \ref{sec_theor}, the numerical simulation corresponding to Sec. \ref{sec_num} with piezoelectric patches attached, and the experimental data. The isotropic damping coefficient used for the mechanical structure in the theoretical model and simulation is $0.001$. Compared to the non-graded metamaterial, no closed form equation exists that governs the bandgap range of graded metamaterials. Therefore, $-20$ dB is regarded as the threshold below which the transmissibility is considered to reflect the bandgap \cite{alshaqaq2020graded}. It can be seen that the three transmissibility curves mutually agree, which validates the theoretical model proposed in Sec. \ref{sec_theor}. The difference of the numerical and theoretical curves versus the experimental curve is mainly due to the plastic cover of the piezoelectric patches, which differs from the homogeneous piezoelectric material property used in theoretical analysis and simulations.

    Additionally, the experimental dispersion relationship of the graded metamaterial is indicated with gray and color circles in Fig. \ref{fig_dispersion}. The experimental wave numbers were obtained by the IWC method with the Fourier transformed time-domain velocity responses of the center-axial points on the host beam under a 30-350 Hz broadband sweep. They indicate good agreement with the numerical dispersion curves. Inside the grading frequency range, the real and imaginary parts of an experimental wave number at each frequency are shown with the same color. These non-zero wave numbers further lead to the experimental amplification factor of the graded metamaterial in Fig. \ref{fig_amp} (a). The experimental amplification factors displayed in Fig. \ref{fig_amp} (a) were calculated by the definition in Sec. \ref{sec_num} with the experimental time-domain velocity responses under harmonic base excitation at the resonance frequencies of different parasitic beams. Compared with the one-pair cases of parasitic beam pairs $L_1$, $L_8$, $L_{13}$, and $L_{15}$ shown as blue dots, the amplification factors of the graded metamaterial are larger, indicating more efficient energy transformation from the host beam to the local resonators under the same input conditions. Fig. \ref{fig_amp} (b) and (c) show an example of the tip velocity responses from the $8^{th}$ pair of parasitic beams and its one-pair case under their resonant frequencies. It can be seen that not only the amplitude $\dot U_{8}$ is larger than its one-pair case. The amplification ratio also outperforms its corresponding one-pair case.

    Figure \ref{fig_power} illustrates a practical example of the harvested power from the graded metamaterial  with the theoretical harvested power curves.
    Under $0.1$ N harmonic input base excitation at the resonance frequency of the $8^{th}$ pair of parasitic beams and its corresponding one-pair case, the experimental harvested power was measured with the output voltage under different loads switched by an electromagnetic relay. It can be seen that all four harvested power curves quickly reach a stable output after a certain load threshold to conduct the bridge rectifier and the freewheeling diode, which shows the load-independent characteristic of the SP-SECE interface circuit. The theoretical harvested power curves indicated with dashed lines agree with the experimental harvested power curves. The difference is mainly due to the loss of the envelope detector. The drop of the experimental power curve of the $8^{th}$ pair of parasitic beams under large load resistance is induced by the parameter loss of the storage capacitor $C_s$ under high voltages. Compared with the one-pair case, the maximum harvested power is enhanced by 489\%. The experimental energy harvesting efficiency is defined as the ratio of the harvested power from the parasitic beams to the base excitation RMS power:
    \begin{equation}
        \eta_{exp}= \frac{P_r}{\frac{1}{T}\int_{T} f_{in}\left(t\right)  \dot{w}_b\left(t\right) \text{d}t},
    \end{equation}
    where $f_{in}$ and $\dot{w}_b$ represent the harmonic input force and the base velocity.  The experimental efficiencies for the $8^{th}$ pair of parasitic beams and its one-pair case are 26\% and 14\%, respectively, which shows the high-capability energy harvesting of the proposed design. They also agree with the numerical results shown in Fig. \ref{fig_efficiency}.
    Broadband energy harvesting can also be naturally realized by varying the different harvesting parasitic beam pairs. Fig. \ref{fig_power} (b) shows the experimental waveform of the piezoelectric voltage $v_p$ and the storage voltage $V_s$ under 1 M$\Omega$ load of the $8^{th}$ pair of parasitic beams and its one-pair case. It can be seen that the piezoelectric voltage of the $8^{th}$ pair parasitic beams is higher due to the wave field amplification from the graded metamaterial, and the storage voltage is also higher.

    \section{Discussion}
    \label{sec_dis}

    \begin{table*}[!t]
    \footnotesize
    \caption {Comparison with Existing Works} \label{compare}
    \centering
   \setlength{\tabcolsep}{1.5mm}{
    \begin{tabular}{llllll}
    \toprule
    \textbf{Reference} & \textbf{Transducer}& \textbf{Resonator size} & \textbf{Operating frequency} & \textbf{Harvested power} & \textbf{Load condition}
    \\ \midrule
       Li et al., 2017 \cite{li2017design}  &  PVDF & 35$\times$35$\times$1 mm$^3$      &   170 Hz       &   0.5 $\mu$W   & 1000 k$\Omega$
    \\
      Chen et al., 2019 \cite{chen2019metamaterial} &   PVDF      &   45$\times$45$\times$2 mm$^3$  &   348 Hz      &  1.25 $\mu$W    &   200 k$\Omega$
    \\
        Hwang and Arrieta, 2018 \cite{hwang2018input}    &   PVDF    &   225$\times$64$\times$0.25 mm$^3$    &   Input independent   & 32 $\mu$W   &  100 k$\Omega$
    \\
        De Ponti et al., 2020 \cite{de2020experimental}    &   PZT    &   (250$\sim$650)$\times$5$\times$5 mm$^3$    &   2000$\sim$5000 Hz   & 70 $\mu$W @ 2050 Hz    &  6.8 k$\Omega$
    \\
        \textbf{This work}   &   PZT      &  $L_r\times$10$\times$2 mm$^3$     &   60$\sim$160 Hz   & 600 $\mu$W @ 91 Hz    &  Load independent
    \\ \bottomrule
    \end{tabular}}
    \end{table*}

    Locally resonant metamaterials fit the scenario of utilizing low-frequency vibrations for self-powered IoT devices. With reference to statistic studies \cite{fan2022applications}, we discuss the PEH performance of the proposed graded-harvester with respect to existing metamaterial-based PEH harvesters as benchmark models. The PEH performance of these harvesters are summarized in Table \ref{compare}, where the chosen transducer type (polyvinylidene difluoride (PVDF) or lead zirconate titanate (PZT)), resonator size, operating frequency, harvested power, and load condition of each harvester are listed in detail. Despite different power performances, all solutions listed provide higher efficiency when compared to conventional energy harvesters because they leverage the local resonant effect. As shown in the table, conventional locally resonant metamaterials (Li et al. \cite{li2017design} and Chen et al. \cite{chen2019metamaterial}) can achieve low-frequency PEH, however, their energy harvesting bandwidth is relatively narrow. By exploiting the nonlinear dynamics or graded designs, the operating frequency\footnote{For multifunction metamaterials, there is no standard definition of the energy harvesting bandwidth. Therefore we use the operating frequency range as a replacement. } can be broadened with multifunction metamaterials, as presented by Hwang and Arrieta \cite{hwang2018input} and De Ponti et al. \cite{de2020experimental}. Building on these pioneering solutions, the graded-harvester presented here not only achieves near-milliwatt power output to accommodate the power demand of low-power consumption IoT devices, but also broadens the operating frequency for low-frequency PEH. Furthermore, its load independence characteristic provides stable output power from heavy to light load conditions. This allows tackling different operational conditions of IoT devices \cite{al2015internet}. Given the simplicity, the robust underlying physics, the inherently broadband and high-capability designs,  the proposed graded-harvester has great potential to power sensors or microcontrollers and realize self-powered IoT devices.

    \section{Conclusion}
    \label{sec_con}
    Based on the idea of graded metamaterials, this paper proposes a graded metamaterial-based energy harvester for broadband and high-capability piezoelectric energy harvesting focusing on ambient vibrations ($<$100 Hz). The broadband energy harvesting target has been inherently satisfied by the spatial-frequency separation of the graded metamaterial design with an efficient grading profile. Furthermore, the high capability energy harvesting target has been investigated by dispersion analysis of low-frequency wave propagation to reveal the slow wave phenomena and the wave field amplification mechanism of the graded metamaterial. By combining the two advantages of the graded metamaterial with the two main goals for piezoelectric energy harvesting, the power performance and the energy harvesting efficiency of the graded-harvester are thoroughly discussed with theoretical, numerical, and experimental analyses. Finally, experiments were carried out to validate the performance of the graded-harvester. It is shown that coupling of the graded metamaterial with the self-powered interface circuit results in a five-fold increase of the harvested power with respect to the conventional harvesting solution commonly adopted in IoT devices. Therefore our proposed design opens up new potential for self-sustainable IoT devices.


\section*{Acknowledgments}
The authors acknowledge support from ETH Research Grant (ETH-02 20-1) and H2020 FET-proactive project  METAVEH under the grant agreement 952039.

\appendix
\section{Equivalent impedance of the SP-SECE interface circuit}

In phase $P_3$ as shown in Fig. \ref{fig_sece}, the $L_i$-$C_p$ resonance should be introduced at the $i_{eq}$ crossing zero point by conducting the switching path with $T_2$. Nevertheless, the cascaded connection of $T_1$ and $T_2$ renders the conduction of $T_2$ dependent on the conduction of $T_1$ to enable a nonzero base current. In order to conduct $T_1$, its base voltage (the bridge-rectified piezoelectric voltage with forward voltage drop $V_F$) should be lower than its emitter voltage $V_{c1}$ by $V_D+V_{BE}$ (the sum of the forward voltage drop of $D_1$ and the base-emitter voltage of $T_1$). In other words, $v_p$ should also have the same voltage drop to conduct $T_1$. Therefore, the phase delay $\varphi$ between the switching instant and current zero-crossing instant is calculated as:
\begin{equation}
    \varphi  = \arccos (1 - \frac{V_D + V_{BE}}{V_{oc}}),
\end{equation}
Assuming that each bias-flip action takes much less time than a vibration cycle, the piezoelectric voltage $v_p$ can be formulated by the following piece-wise equation:
    \begin{equation}
    \small
	\begin{aligned}
	&{v_p}(t)=V_{oc}\times&
	\left\{
	\begin{aligned}
	&\cos \varphi - \cos (\omega t),&\varphi &\le \omega t<\pi  + \varphi ;\\
	-&\cos \varphi  -\cos (\omega t),&\pi  + \varphi &\le \omega t< 2\pi  + \varphi. \\
	\end{aligned}
	\right.
	\end{aligned}
	\label{eq:spsece}
	\end{equation}

Through a Fourier analysis, the fundamental harmonic of $v_p$ can be derived based on the piecewise expressions in \eqref{eq:spsece}. The fundamental component of $v_p$ is denoted as $v_{p,f}$ \cite{Liang2012}. Based on harmonic analysis, the dynamics of SP-SECE can be formulated in terms of equivalent impedance, which is obtained in the frequency domain as follows:
    \begin{equation}
	Z_{e} = \frac{1}{\omega C_p}\left[ \frac{2}{\pi }\left( 1+\cos 2\varphi  - j\sin 2\varphi \right) -j\right].
	\label{eq:Zespsece}
	\end{equation}

 \section{Displacement amplitude of the parasitic resonators}

	In order to solve the governing equations representing the graded-harvester, the solution of the host beam deflection is obtained with the separation of variables as $w\left( {x,t} \right) = \sum\limits_{n = 1}^N {{\phi _n}(x){\eta _n}(t)} $, where $\phi _n$ and $\eta _n$ represent the mode shape and modal coordinate of the $n^{th}$ mode, respectively. By substituting this latter into Eq. \ref{eq:hostbeam}, applying the orthogonality relationships for $\phi _m$, and assuming harmonic excitation and solutions \cite{sugino2016mechanism}, the relative displacement amplitude of $r^{th}$ pair of parasitic resonators can be expressed as:
    \begin{equation}
        {U_{r}} = \frac{{{\omega ^2}{W_{b}} + {\omega ^2}\sum\limits_{n = 1}^N {{H_{n}}{\phi _n}\left( {{x_r}} \right)} }}{{\left(1+ K_{er}/k_{mr} \right)\omega _r^2 + j2\left( 1 + D_{er}/D_{mr} \right){\zeta _r}{\omega _r}\omega  - {\omega ^2}}},
        \label{eq:Ur}
    \end{equation}
    where $W_{b}$ and $H_{n}$ represent the amplitude of the harmonic base excitation and the $n^{th}$ mode modal amplitude, respectively.

    Then, by plugging Eq. \ref{eq:Ur} into the modal form of Eq. \ref{eq:hostbeam}, the governing equation can be simplified as:
    \begin{equation}
    \small
    \left(\omega_{m}^{2}-\omega^{2}\right) \frac{H_{m}}{mL}-\omega^{2} \sum_{n=1}^{N} \sum_{r=1}^{R} \hat{M}_{r} C_{r} \phi_{m}\left(x_{r}\right) \phi_{n}\left(x_{r}\right) H_{n}=Q_{m}
     \label{eq:simpeq}
    \end{equation}
    where ${\hat M}_r=2M_r/\left(mL\right)$ is the normalized equivalent mass of $r^{th}$ pair of parasitic resonators. The coefficient $C_r$ and the modal force item $Q_m$ can be expressed as:
    \begin{equation}
    \small
    \begin{aligned}
    C_{r} &=1 + \frac{\omega^{2}}{\left(1+ K_{er}/k_{mr} \right) \omega_{r}^{2}+j 2\left( 1 + D_{er}/D_{mr} \right) \zeta_{r} \omega_{r} \omega-\omega^{2}}, \\
    Q_{m} &=W_{b} \omega^{2}\left(\sum_{r=1}^{R} \hat{M}_{r} C_{r} \phi_{m}\left(x_{r}\right)+\frac{1}{L} \int_{0}^{L} \phi_{m}(x) d x\right) .
    \end{aligned}
    \end{equation}
    By writing Eq. \ref{eq:simpeq} into matrix form and accounting for the contributions of all $N$ modes together,  the modal amplitude $H_{n}$ can be solved by matrix inversion. Finally, by substituting $H_{n}$ back into Eq. \ref{eq:Ur}, $U_{r}$ can be solved.

 \bibliographystyle{elsarticle-num}
 \bibliography{ref}





\end{document}